# iVRNote: Design, Creation and Evaluation of an Interactive Note-Taking Interface for Study and Reflection in VR Learning Environments

Yi-Ting Chen[1][*]   Chi-Hsuan Hsu[1][†]   Chih-Han Chung[1][‡]   Yu-Shuen Wang[1][§]   Sabarish V. Babu[2][¶]

[1]National Chiao Tung University    [2]Clemson University

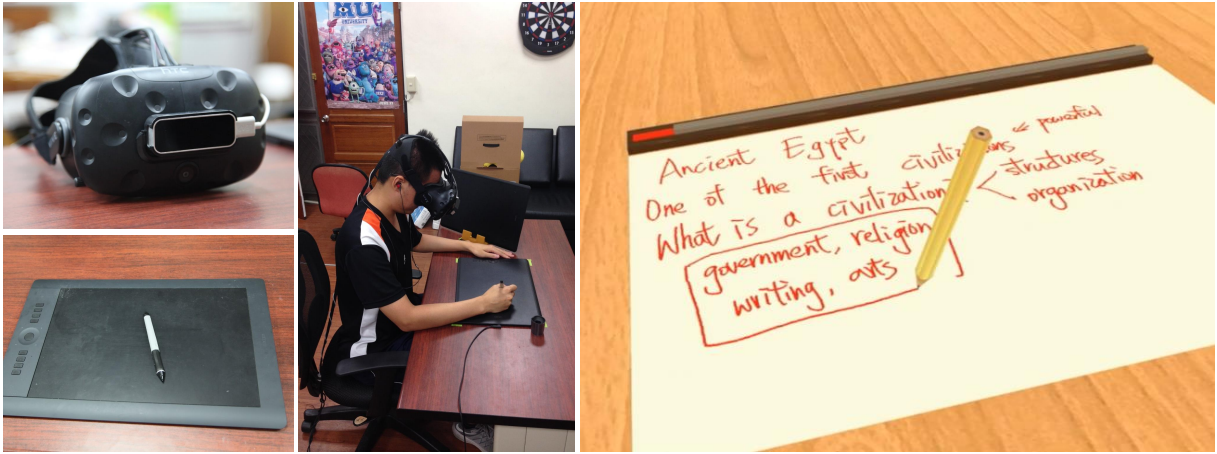

Figure 1: We present iVRNote, an interactive note taking interface, to help students study video lectures in VR (middle and right). To enable students to take notes in VR (right), we install a digital tablet (bottom left) on the desk and track a stylus's position and orientation using the tablet's interface and the leap motion controller (top left).

**ABSTRACT**

In this contribution, we design, implement and evaluate the pedagogical benefits of a novel interactive note taking interface (iVRNote) in VR for the purpose of learning and reflection lectures. In future VR learning environments, students would have challenges in taking notes when they wear a head mounted display (HMD). To solve this problem, we installed a digital tablet on the desk and provided several tools in VR to facilitate the learning experience. Specifically, we track the stylus' position and orientation in the physical world and then render a virtual stylus in VR. In other words, when students see a virtual stylus somewhere on the desk, they can reach out with their hand for the physical stylus. The information provided will also enable them to know where they will draw or write before the stylus touches the tablet. Since the presented iVRNote featuring our note taking system is a digital environment, we also enable students save efforts in taking extensive notes by providing several functions, such as post-editing and picture taking, so that they can pay more attention to lectures in VR. We also record the time of each stroke on the note to help students review a lecture. They can select a part of their note to revisit the corresponding segment in a virtual online lecture. Figures and the accompanying video demonstrate the feasibility of the presented iVRNote system. To evaluate the system, we conducted a user study with 20 participants to assess the preference and pedagogical benefits of the iVRNote interface. The feedback provided by the participants were overall positive and indicated that the iVRNote interface could be potentially effective in VR learning experiences.

## 1 INTRODUCTION

Virtual Reality (VR) learning has been promoted for many years worldwide, which has several advantages such as low cost, comfort, and free of distance limitation [35, 40, 48]. As VR technology become more accessible and ubiquitous, students who are willing to learn have the potential to take any course they want without moving to a completely unknown city or country by participating in the VR learning experience remotely. It has been shown that although thousands of students participate in online courses in the physical world, the completion rate of most such courses is below 13% [34]. One of the challenges noted is that in online situations, students are often distracted by external sources, such as television, social media or mobile phone applications. However, VR has the potentially to immerse and isolate learners in the classroom experience, and facilitate greater concentration and attention to the learning materials, as preliminary evidence shows in the use of VR learning experiences for students with ADD [2, 36]. They also can potentially learn anytime and anywhere, pause the lecture to take notes, and revisit the lecture to understand difficult materials. Although the technological potential for such widespread VR classroom experience exists, there is a lack of tools and interaction metaphors in a VR classroom settings to facilitate active immersive note taking in concert with lecture viewing, and tools for review and reflection. Current best existing interactive note taking and learning interfaces in VR do not have the complexity to facilitate active learning experiences beyond merely the capability of writing [6, 37], and these tools and techniques have not been empirically evaluated with regards to their pedagogical benefits, preference and impressions. Therefore, in order for future VR classroom experiences to be pedagogically effective, efficient and engaging, the interactive features, similar to a physical world classroom environment, must be designed, developed and evaluated for active learning experiences in VR. Thus, the above mentioned problem space motivated us in this contribution to design, create and evaluate the iVRNote interaction metaphor for interactive note-taking, study and reflection in VR classroom situations (see Figure 1).

[*]e-mail: asd2846531@gmail.com
[†]e-mail: ace810408@gmail.com
[‡]e-mail: chihhichihhi@gmail.com
[§]e-mail: yushuen@cs.nctu.edu.tw
[¶]e-mail: sbabu@clemson.edu

Leveraging the capabilities of commercially available VR systems, we aimed to help students immerse and engage with classroom materials via interactions with our iVRNote in a custom build collaborative VR classroom environment that served as a backdrop for the development of the interactive features and evaluation. Considering that note taking is an important feature [31] for students towards active learning of any new material, and the inspiration from a handwriting interface in VR [38], we implemented iVRNote in the VR classroom. Specifically, we installed a digital tablet on a desk for users to write and draw. The handwriting and drawing trajectories were recorded and displayed on a note paper in the VR classroom. To match the visual and haptic feedback perceived by users, we tracked the physical stylus and applied the tracked position and orientation to render the virtual stylus in the student's virtual hand that was co-located with the physical stylus. The tracking can be done by the tablet itself if the stylus is close to the tablet. However, if this is not the case, we track the stylus according to stereo infrared images captured by a leap motion controller mounted on the HMD. The two tracking results switch over based on the presence and absence of signals received from the digital tablet.

Several physical limitations can be eliminated in VR because the whole learning process via iVRNote has been digitized. For example, students may need to correct or post-edit their note taking when taking classes. Digital note taking systems have great advantages over traditional note taking in this situation. iVRNote provides users with not only basic functions, but also advanced operations such as post-editing of notes and review by selection. The post-editing could be very useful when students attempt to insert a line of text into the middle of a paragraph. In a traditional way, they have to write the text next to the paragraph and draw an arrow to point out the position they would like to insert. While using iVRNote, they can move the pre-written text to accommodate the space for insertion, which would make the notes well-organized. In addition, since traditional notes are taken by pen and paper, such notes can be read only in a review phase. Students could be confused if the notes are too brief or contain mistakes. In contrast, when students take notes in using iVRNote, the time of every stroke is recorded by the system. Hence, in the review phase, if students do not understand a part of the note and fail to build their own knowledge, they can specify in the digital note where they are unclear, and then watch the corresponding part of the video lecture again.

To evaluate iVRNote interaction metaphor's effectiveness, we conducted a user study with 20 participants. They were asked to study online video lectures in VR and in the physical world, and compared the behaviors of note taking in these two platforms. Experimental results indicate that the presented VR note taking system can help students study and review lectures. The participants also gave us quite positive feedback and confirmed the feasibility of effectively using iVRNote in virtual classroom learning situations.

## 2 RELATED WORK

### 2.1 Online Learning Interactions

Online learning has becoming popular. An important advantage is that students can study a lecture anywhere and anytime as long as they can access the Internet [33]. As many online learning systems are built using video lectures, students can pause the video to think and reflect about the materials, or replay the video to revisit if they have problems in interpretation of the material. There were many studies that have provided systematic analyses [7, 29, 32, 41] to prove the effectiveness of online learning environments. However, on the negative side, there were also studies showing that students pay less time to take online courses, as compared to students who take traditional classes [10]. The situation is even worse for male students, younger students, African-American students, and low-achieving students [46]. Moreover, there were many studies presented to discuss the high dropout rates of massive open online courses (MOOCs) [34]. Several negative reasons for the high dropout rates include the lack of time [8], course difficulty, and lack of support [28].

### 2.2 Note Taking in Education

Although many researches have shown that note taking can facilitate study [44], taking notes longhand and using laptops can be considerably different. Specifically, because most students can type significantly faster than they can write, they tend to transcribe content mindlessly or indiscriminately [11]. On the other hand, when students take notes longhand, they have to select important information to include in their notes, which forces them to process content deeply in the class [23]. Accordingly, note taking longhand has more advantages in improving learning and retention, whereas the use of laptop can leave a more complete record for future review [31]. Given the different ways of note taking, the studies in [31] show that participants who took notes longhand studied much better than those who used laptops. In addition, they found that telling students who used laptops not to take notes verbatim could not prevent the behavior. They also found that, even when participants could review their notes before the test, the experiment results were the same.

### 2.3 VR in Education

There have been many VR applications presented for training skills. For example, surgeons can learn and practice surgical skills in a virtual operating room without any danger to patients [21, 22]; pilots can practice takeoff and landing skills in a virtual airplane without causing death and financial loss [4, 47]; and ADHD-diagnosed children can train themselves to rehabilitate their attention deficits [40]. In addition, VR can be used to develop spatial abilities. Kaufmann et al. [25] presented a software called *Construct3D* for students at high school and university level to create 3D models by a two-handed 3D interaction tool. Anecdotal evidence in their pilot study shows that Construct3D can help students learn and encourage experimentation with geometric constructions. Angulo and Velasco showed that VR can support the design of architectural spatial experiences [3]. The real-time feedback in VR environment greatly improves student's spatial design and enhances results. Abulrub et al. demonstrated that, by using 3D interactive tools to examine 3D prototypes in VR, engineering graduates can apply theoretical knowledge to real industrial problems and obtain practical experience [1].

The survey on existing solutions to virtual learning environment suggests that providing learners with a high level of immersion and involvement can facilitate learning [39]. Chittaro and Buttussi [14] educated passengers by allowing them to experience a serious aircraft emergency with the goal of survival. Experimental results showed that VR educated participants can retain the safety procedures over a long time span. Cheng et al. [13] attempted to explore whether a virtual foreign environment can help people stay engaged in learning language and culture. They adapted a 3D video game called *crystallize* to a VR version and let users interact through verbal and physical motions. Bailenson et al. [5] designed several experiments in a VR classroom for the study of human behavior and communication. From the experiments, they found that students' learning performance can be effected by their locations relative to the teacher's field of view and other model students or distracting students. Since there are many VR educational applications to be listed in the paper, we refer readers to [12, 19, 30, 39] for details.

### 2.4 Note Taking in VR

Although, note taking can be facilitated via keyboard based text entry, and illustrations can be sketched via mouse based interaction, these interaction metaphors in the past could not be readily afforded in immersive VR situations due to challenges in tracking and rendering external objects as well as our corresponding fine motor actions on them [26]. Natural interaction metaphors have been proposed for note recording via a tracked stylus and tablet surfaces in VR to allow users to have visuo-haptic feedback, and perform the fine motor task of recording observations in an intuitive manner. Thus, leveraging the learned perceptual-motor coordination of the same task from the physical world writing, as well as the tactile augmentation between the stylus tip and virtual canvas [17, 37]. These natural metaphors have also been shown to promote a greater sense of presence, and improve the efficacy of task performance in VR. Many studied have investigated the effects of a stylus or pen based interaction to mouse based interaction in digital input on horizontal interfaces [9, 18]. Research shows that participants preferred a pen based interaction on a digital plane, as compared to a mouse based interaction [16]. The design and development of natural pen or stylus based interaction in VR in the past suffered from latency issues between the motor component and corresponding rendering of the content on the canvas, which was detrimental to participant performance in writing and drawing [37]. However, contemporary VR simulations afford high resolution viewing, high frame-rate tracking and rendering, such that using hybrid tracking solutions a natural pen based interactions for writing and sketching can be readily implemented. This has led to the popularity of writing application in contemporary VR systems, such as Google's Tilt Brush [24].

Although VR sketching and writing by itself have been formulated and tested with regards to variables such as presence and task performance, there is a need for more advanced 3D interaction metaphors associated with note taking and sketching towards enhancing learning outcomes, engagement, study and reflection in VR classroom situations. This contribution aims to design, create and evaluate these advanced VR interaction metaphors (via iVRNote) for usable, engaging, and effective pedagogical interaction in VR classroom experiences.

## 3 REQUIREMENT ANALYSIS FOR IVRNOTE

In this research, the goal was to help students study in VR as they are used to studying in the physical world. Therefore, similar to a traditional setting, in VR, students can also sit in front of a desk and could watch the video lecture to study. Considering that note taking provides several benefits, such as improving memory, aiding in the organization of information in a class presentation, students should be able to take notes in VR as well. To achieve an easy-to-use VR education application, we discussed with several cohorts of undergraduate and graduate students in our University, as well as with a group of education experts, and arrived at several system requirements that should be included. They were as follows:

1. *Support when students have problems in understanding materials.* Referring to external resources, such as Wikipedia, for understanding prerequisites that are not explained in the lecture video is important in study. This feature can be easily achieved in the physical world. However, taking off the HMD and leaving the VR classroom is tedious and would prevent students from reaching external resources when they have problems in understanding lectures. Students should have ways to obtain solutions in VR.

2. *Visibility of stylus.* Students would be blind to the physical environment when they wear an HMD. Therefore, showing

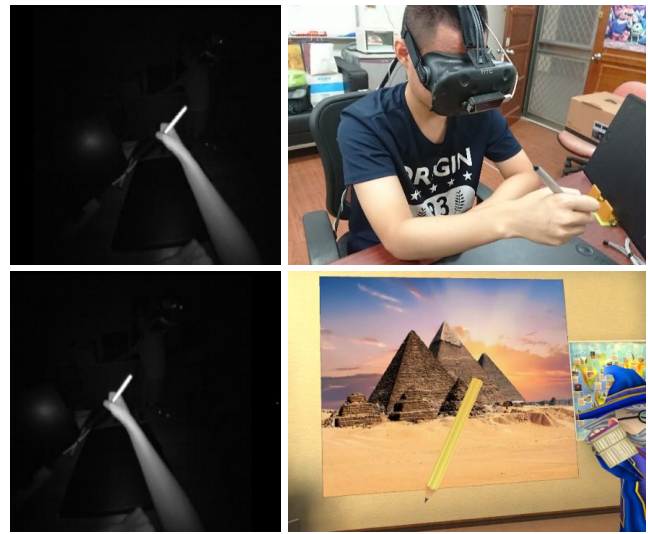

Figure 2: (Left) Stereo infrared images captured by the leap motion controller. (Top right) The photograph shows how the user holds the physical stylus. (Bottom right) We track the stylus' position and orientation in the physical environment and shows the virtual stylus in the VR classroom.

a virtual stylus as it appears in the physical environment is important. Students should be able to reach out their hand for the stylus, when it is somewhere on the desk, and they also should be aware of (visually perceive) where the virtual stylus could touch the note paper before they write.

3. *As many words can be written on a note paper.* Students may need to write many words and draw diagrams on a note paper to clearly describe and illustrate an important idea. In other words, the note taking interface should be of high precision so that students can write words in the VR classroom as small as they write in the physical environment (similar in writing quality to the physical world).

4. *Correction and post-editing of notes.* Note taking inevitably can result in mistakes. Students should be able to erase the notes when they write something incorrectly. It would be great if students can post-edit and organize the notes when they would like to re-organize the notes after the class. This characteristic is important because students have to take notes and watch a video lecture simultaneously. Students typically may not have much time to organize their notes during a class lecture.

5. *An efficient way to take notes.* There can be many important points to be written in a class. While students have to watch a video lecture, think about the materials, and take notes simultaneously, they can be very busy in doing everything well. In addition, when the lecture goes to the next slide, students may not have enough time to write down what they would like to record. Hence, an efficient way to take notes is needed.

## 4 IVRNOTE SYSTEM IMPLEMENTATION

The presented note taking system aims to help students study in VR. Therefore, we consider the requirements raised by the students and educators to implement this system. Specifically, we build a 3D classroom, setup a set of corresponding hardware, integrate these virtual and physical objects seamlessly, and then provided a framework for the iVRNote interaction metaphor.

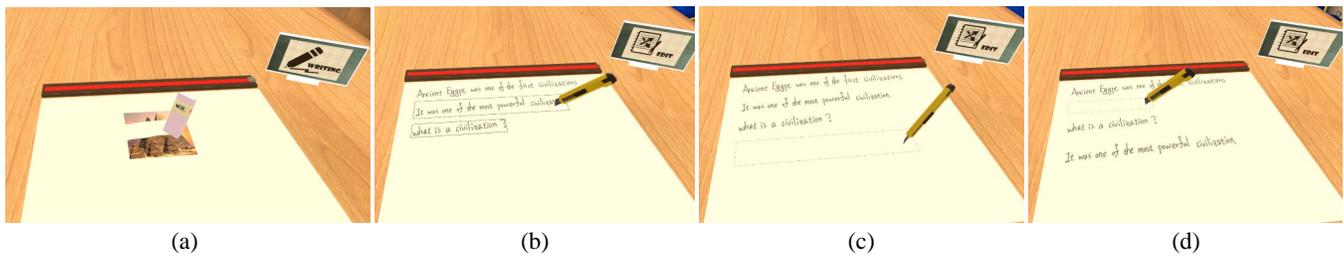

Figure 3: Basic functions implemented in the presented iVRNote. (a) Eraser. (b-d) Post-editing of notes. Students can select a part of the note and then move it to different positions.

### 4.1 Hardware and Software

We implemented the VR classroom by integrating an HTC Vive, Leap motion controller, and a Wacom tablet ( Intuos Pro large, PTH-860/K0). Figure 1 shows the hardwares used in the presented system. Specifically, we setup a desk in the physical environment and placed a Wacom tablet on the desk for users to take notes. Similarly, we set a desk and a note paper in the virtual environment. The 1:1 correspondence between virtual and physical objects were determined by the HTC tracking in advance. In addition, we installed the leap motion controller in front of the HMD to track the physical stylus if the stylus is far away from the tablet. Users would be able to see a virtual stylus as long as the physical stylus is within the field of view (**R2**).

### 4.2 Stylus Tracking

Our system tracks the physical stylus and applies the tracked position and orientation to render the virtual stylus in the VR classroom. Thanks to the electromagnetic sensors installed in the Wacom digital tablet, when the stylus is close enough (roughly 1 cm) from the tablet, it's position and orientation can be obtained via the Wacom API[1]. When the stylus is far away from the tablet, we track the pose of the stylus using a computer vision technique. The two versions of the stylus tracking are switched according to the sensor readings obtained from the API.

We installed a leap motion controller at the front of the HMD to capture stereo infrared images (Figure 2). To ensure that the stylus is salient in the infrared spectra, we wrapped the stylus in a segment of IR reflective tape. The stylus in the infrared images would be the brightest after the illumination (by the leap motion controller). Hence, we identify the pixels that have intensities larger than 250 in the stereo infrared images, compute their correspondence using the iterative closest points method [42], and then obtain the horizontal disparities between the corresponding pixels. Since the horizontal disparities can be used to determine the depth from the HMD (i.e., user's head) to the stylus, we un-project the stylus pixels from 2D back to 3D. In addition, limited by the stylus' shape, these un-projected 3D pixels were expected to appear as a thin cylinder. Subsequently, we applied principal component analysis (PCA) [15] to compute the main axis of these 3D pixels, and then determine the stylus' position and orientation. We also applied the Kalman filter [45] to smooth the axis as illumination and occlusion problems may occur with the user's hand during stylus tracking.

We wrapped the stylus in a segment of IR reflective tap to facilitate stylus tracking. Although the tape could be occluded by the user's hand, we found that the approach worked well in the experiment because the stylus was on the hand web when participant was writing or drawing. A large portion of the tape was still visible in this situation, no matter participants held the stylus near or far away from the nib. However, lost tracking to the stylus can still occur if participants shook or rotated their hands arbitrarily, as the tape was mostly occluded. Regarding the two versions of the stylus' pose, we adopted the result obtained from the Wacom API if the stylus can be tracked by the device, which was indicated by a parameter received from the API. If this was not the case, we adopted the result tracked by the computer vision technique. The switch may induce motion discontinuity because of the imperfect tracking results. We interpolated the two poses in our implementation to ease the visual artifacts.

## 5 STUDYING IN VR USING IVRNOTE

We build a VR classroom for students to study video lectures that are publicly available online. To engage students' attention in the class, we create a 3D classroom and placed a projection screen at the front to display the lecture videos. Students sit in front of a desk as they usually do in the physical classroom environment. In addition, they take notes on the electronic papers that are placed on a virtual desk (**R3**). Considering that a digital tablet can detect finger touch as well, we let users flip the note to previous and next pages by swiping their finger left and rightward on the tablet surface. To make this interaction metaphor intuitive, we implemented the 3D page flipping animation. The animation is predefined and would play automatically when the finger swiping event is detected.

In addition to taking notes by using a virtual stylus, students can switch the stylus to other tools, such as utility knife, magic stick, and markers, in VR. To achieve this, they simply click the forward and back buttons on the physical stylus. Figure 4 shows the tools available in the presented system. Each tool corresponds to a function that will be described in the later sections. We also show a small card next to the note paper to indicate to users the current function if they are new to the system. Note that we encourage readers to view the accompanying video for the introduction to these tools as the interactive operations are difficult to visualize in still images. If participants were using the VR classroom system for the first time, then they were provided with an interactive training session to acclimate themselves to these 3DUI educational interaction metaphors in a manner similar to best existing VR applications [26].

### 5.1 Eraser and Post-Editing

Mistakes often occur when students take notes. Therefore, considering that a stylus in iVRNote has two sides, students can flip the stylus to become an eraser (Figure 4 (b)) and erase the notes if they make mistakes (Figure 3 (a)). In addition to this fundamental function, the presented VR classroom allows users to post-edit the notes, which would be very useful when they attempt to insert a line of text into the middle of a paragraph. In a traditional manner, users have to write the text next to the paragraph and draw an arrow to point out the position they would like to insert a given text. While in the VR classroom, by switching the stylus to the utility knife (Figure 4 (f)), they can cut the pre-written text and virtually move or translate it to a different position to accommodate space for insertion (Figure 3 (b-d)). The post-editing helps students organize the notes well (**R4**).

---
[1] https://developer-docs.wacom.com/pages/viewpage.action?pageId=10422351

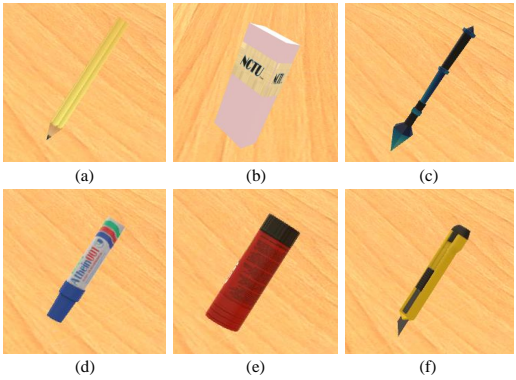

Figure 4: Students can use different tools in the iVRNote to help them learn lectures, take notes, and review. Each tool corresponds to a function. (a) Stylus. (b) Eraser. (c) Magic stick. (d) Marker. (e) Glue. (f) Utility knife.

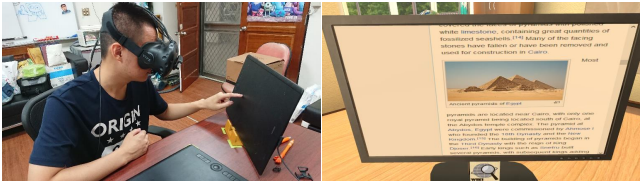

Figure 5: Students can check the Wikipedia page on a display on their desks in the virtual classroom, if they encounter difficult words and terms in a lecture. (Left) The student scrolls the web page by swiping his finger on the virtual display that is physically co-located with a Wacom tablet for passive haptic feedback. (Right) The Wikipedia web page is shown in the VR classroom on a virtual desktop computer.

### 5.2 Virtual Wikipedia Reference for Learning

Students may encounter difficult words or terms in classes. Considering that the instructor cannot answer questions in the presented VR classroom, we let students check Wikipedia for obtaining solutions (**R1**). To implement this idea in iVRNote, we placed a virtual screen on the left of the desk that shows the Wikipedia page, which was co-located with a physical display for passive haptic feedback. Students can examine the screen and then speak the word they would like to search. For example, when they say *"search pyramid"*, the presented system would process the voice command by the Google speech API and then link the page to "pyramids." In addition, since the page can be too long to be fully displayed on the screen, students may need to scroll down the page in order to read the whole description. On the virtual display, students can move pages up and down by swiping their fingers on the virtual display surface. Figure 5 shows the operation.

### 5.3 Taking Notes by Virtual Photography

In recent years, many students take notes by taking pictures to obtain slides or writings on the blackboard so as to save effort and pay more attention to listening to instructors. These pictures are then used in the review. However, when students are in a class, adding comments and annotations to the pictures is problematic because cameras and notes are independent of each other. Since both pictures and notes in the VR classroom are digital, the problem can be easily solved. That is, in iVRNote, users first take a picture with a predefined gesture, and then sketch an area on their note for embedding the picture (**R5**). Figure 6 shows the steps. To take a picture, users pinch the thumbs and the forefingers of their two hands to define a rectangle in 3D. The presented system obtains the capturing area on the slide by ray tracing from the user's head through the user-defined 3D rectangle. The four corners of the capturing area are then rendered on the slide for indication. Students can still fine tune the area by moving their pinched fingers. The picture would be captured when they unpinch the fingers. To embed the picture, students have to sketch an area on the note paper by using the glue (Figure 4 (e)). A bounding box is then computed for embedding. The system automatically pauses the class if pinch gestures are detected and continues the class after the pictures are embedded.

It is suggested that users can drop the stylus on the table when taking a picture, and then pick up the stylus to sketch an area on their note for embedding the picture. Otherwise, the stylus can interfere with the gesture detection by the leap motion controller.

### 5.4 Revisiting a Specific Part of a Lecture

In the VR classroom, with iVRNote students can revisit the video lecture by dragging the time slider. To achieve this, they have to switch the stylus to a magic stick (Figure 4 (c)) and drag horizontally on the tablet. The time slider on the top of paper shows the lecture time. Considering the small field of view of the HMD, we render a preview above the time slider for students to discover the slides (Figure 7) so that they can roughly obtain the relations between the lecture time and the content.

### 5.5 Review by Selecting Notes

Studying notes is an effective way to review a lecture. However, notes can be too brief to understand or contain errors because students have to listen to instructors and write important points simultaneously. As a result, students may have difficult interpreting their notes or obtain incorrect knowledge from a review. In order to address this difficulty, we record the time of each stroke written by students so that they can review the video lecture if they get confused from their notes. In the process, they can select a part of the notes by using a marker that is shown by the system at the time when the notes were written by the student (Figure 4 (d)). Then, our system would move the lecture video to the time corresponding to the selected note (Figure 7) and review the lecture.

### 5.6 Hardware Specifications of the System

We have implemented the VR classroom and iVRNote interactions on Unity 3D and ran the program on a desktop PC with Core i7 3.0 GHz CPU and NVIDIA GeForce GTX 970 graphics card. The system runs in real time at 70Hz, so that users can interact with the virtual environment smoothly.

## 6 USER STUDY

We conducted a user study with 20 participants to evaluate the user experience, user impressions and effectiveness of note taking and reviewing in VR using iVRNote and in the physical world (PW). The participants were asked to study online lectures in both conditions.

### 6.1 Physical World Condition

We compared VR to PW condition because we aimed to know whether transforming students to a VR classroom with the iVRNote interaction metaphor can help them engage with online lectures and study better. We also aimed to understand whether the presented note taking system iVRNote can help students learn and review online lectures effectively. Hence, in an experiment, we let participants in the PW group used physical pens and papers to take notes. To allow participants in this group to take notes in their usual way as they do in a traditional in-class session, we did not setup the VR and PW conditions to be highly similar, except that the size of the physical and virtual notepads were the same. Specifically, participants in PW condition sat in front of a desk and studied by using a personal computer, whereas participants in VR condition watched the lecture video on a projector screen in the virtual world. The notepad in the VR condition was fixed at a location on the desk, whereas the paper in the PW condition could be moved. In addition, the pen used in the PW condition was thinner and lighter than the Wacom stylus, and the ink stroke size in PW condition was thinner than that in VR condition. We were aware of the influence of pen parameters on writing [20]. However, the constant size and the weight of the stylus, and the ink stroke size were considered to be a limitation in the current VR condition, although they were not ergonomically challenging to use.

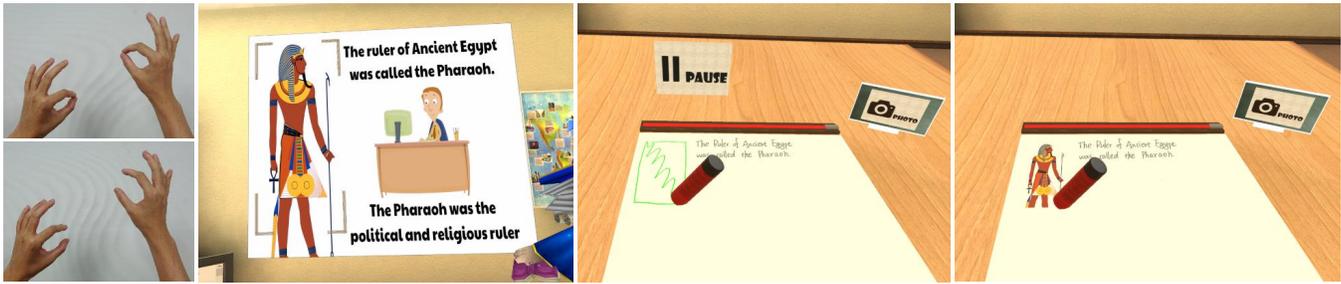

Figure 6: Students can take notes by photograph in the VR classroom. (Left) The pre-defined pinch and un-pinch gestures used to take photograph. (Middle left) Students can move their hands with thumbs and forefingers pinched to define the cropping area. (Middle right and right) After they take the picture by un-pinching the fingers, they draw an area on the note paper to paste the picture.

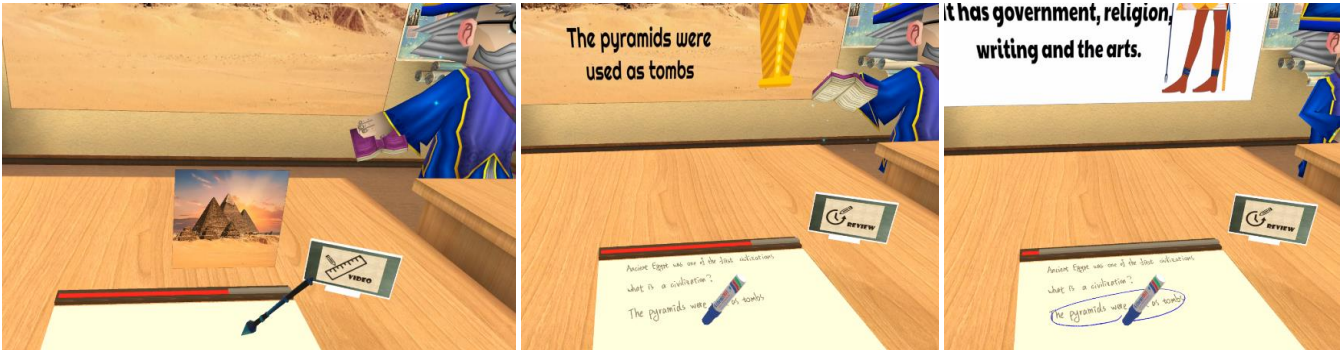

Figure 7: (Left) Students can revisit a part of the video lecture by dragging the time slider. (Middle and right) The presented VR classroom records the time of each written stroke. If students have problems in interpreting their note, they can simply select the confusing part. Our system will replay the video lecture from the time stamp corresponding to the selected note for students to revisit.

### 6.2 Participants

Participants were recruited via an online recruiting system, and came from diverse backgrounds and majors. There were 6 female participants and 14 male participants. Their ages ranged from 21 to 27 (M=23.4 and SD=1.88); and 95% of them had experience of taking video lectures in remote learning systems. In addition, participants were interviewed to insure that they were unfamiliar with the lecture topics of the study.

### 6.3 Study Design

In a within-subjects repeated measures evaluation, participants were randomly assigned to one of two conditions in the first session and to the other condition in the second session. Participants were also randomly assigned to learn either a Meteorology (Science topic) or History (Humanities topic) related topic in the first and second sessions. These two topics are not something one would know through general knowledge, or could reason by common sense. There were advanced topics that needed to be learned through study and reflection. We also confirmed by interviewing the participant that they were not familiar with the topics before the study. The order of learning condition and topic was counter-balanced between the two sessions. The research question was, *"to what extent is the VR classroom with iVRNote effective, user friendly, engaging and satisfying, as compared to a physical learning environment?"* Our study aimed to compare and contrast, and verify how the iVRNote interaction metaphor for note taking, study and reflection in a VR classroom differed from a traditional pen and paper method for the same purpose.

### 6.4 Procedure

There were four phases in each session. The first two phases and the last two phases were three days apart. First, in the *training phase*, the participants learned how to use the assigned system. After being familiar with the system, the participants learned the lecture topic using the assigned system in a *learning phase*. During the learning phase, participants were asked to take notes like they normally do in a traditional classroom setting. In addition, they were free to pause and continue the lecture if they needed some time to take notes. The length of the learning phase was approximately 45 minutes. This phase ended after the participants learned the lecture topic. After three days or more, the participants came back and were asked to review the lecture for a specified period of time by revisiting the lectures, notes, and review features in the same system in a *reviewing phase* - enabling researchers to evaluate the review components of the system. Participants were told that they would be tested on the topic immediately after the review session. After the review phase, participants were then given a 10 question test on the lecture topic in a *testing phase*.

At the conclusion of the testing phase, participants then completed a post-experiment questionnaires consisting of usability, ease of use, presence [43], and learnability of the systems that covered both usability and user experience metrics. After the conclusion of the post-experiment phase, participants were scheduled for the second session. In the second session, participants were randomly assigned to the other learning system and were asked to learn a different topic as compared to the first session. Then, the training, learning, reviewing and testing phases were repeated using the other system with the second lecture topic. At the end of the second session, participants were again given the post-experiment questionnaire, a preference questionnaire and were interviewed, and then were finally debriefed prior to dispensing the incentive for participation and concluding the study.

## 6.5 Quantitative Results

### 6.5.1 Test Performance

The participants' post-experiment session test scores in PW and VR were subjected to a 2x2 repeated measures ANOVA analysis. One within subjects factor was the topic of study (Metereology vs. History), and the other within subjects independent factor was the condition (VR vs. PW). There was no significant main or interaction effects in the participants test score performance. In order to test if the learning performance was equivalent between the VR and PW conditions in each of the topics (Humanities or Science), we conducted an equivalence test between the mean scores in each topic to demonstrate that the means between VR and PW were equivalent. Equivalence is demonstrated by showing that the 95% confidence interval for the difference between the scores is entirely within a range of -25 and +25, by running two one-sided tests (TOST). The equivalence test between the mean scores in the VR (M=69.0%, SD=14.45) and PW (M=75%, SD=23.34) conditions in the Climate topic were found to be statistically equivalent, Lower 95% CL=-20.27 $p$=0.016 and Upper 95% CL = 8.28 $p$=0.0006. Likewise, the equivalence test between the mean scores in the VR (M=71.25%, SD=16.70) and PW (M=79.75%, SD=12.17) conditions in the History topic were found to be statistically equivalent, Lower 95% CL=-19.24 $p$=0.0077 and Upper 95% CL = 2.25 $p$=0.0001.

### 6.5.2 Preference Questionnaire

After experiencing either the VR or the PW condition in either session, we asked participant their impression of the learning system. The questions in the survey addressed the following dimensions (based on the IBM System Usability questionnaire [27]) to what extent the learning system motivated the participant to learn, allowed participants to focus or attend to the study, enabled the participants to overcome learning difficulties via its features, made participants willing to use the system, user satisfaction and usability. Participants rated each dimension on a 5-point Likert scale with higher scores indicating higher preference for the system. A Wilcoxon's Signed-Ranks test was administered on the scores on each dimension of the preference questionnaire between the VR and PW conditions revealed the following. The analysis (Figure 8) revealed that participants rated the VR condition (M=3.85, SD=0.81) significantly more motivating to learn than the PW condition (M=3.40, SD=0.68), Z=117.0, $p$=0.049. Participants rated the VR condition (M=3.80, SD=0.83) significantly higher in enabling them to focus or attend to the lecture and study than the PW condition (M=3.10, SD=0.78), Z=124.0, $p$=0.018. Participants rated the VR condition (M=3.60, SD=0.82) significantly higher in enabling them to overcome learning or study difficulties than the PW condition (M=3.10, SD=1.02), Z=74.5, $p$=0.032. Finally, participants rated the VR condition (M=3.85, SD=0.74) significantly more satisfying overall than the PW condition (M=3.35, SD=0.74), Z=114.5, $p$=0.047.

### 6.5.3 iVRNote Writing Performance

We compared the size of characters written in PW and VR because the participants were asked to take notes in the virtual and the physical environments. Specifically, for each participant, we manually cropped the characters written by using the virtual tablet surface, and the traditional pen and paper, respectively, and then computed the diagonal length of each character. Note that we measured the size of each character by its diagonal length rather than the x-height because the notes were in Chinese. In addition, the characters in both PW and VR were in the same measure, and the size and location of the canvas/paper were the same in both conditions. A Wilcoxon's Signed-Ranks test revealed that participants wrote with character sizes that were significantly larger in the VR condition (M=3.20 cm, SD=1.15 cm) as compared to the PW (M=1.01 cm, SD=0.71 cm), Z=-3.82, $p$<0.001, Figure 9. There could be two reasons that lead to this result. First, we set the stroke to be thicker in the VR classroom to reduce flickering artifacts caused by the low resolution of the HMD. Second, the participants were unfamiliar with taking notes in the virtual environment. The note with the maximum character size was written by one participant (6.48 cm) in the VR condition. We found that the participant's note taking in the VR classroom was somewhat haphazard, and was thus an outlier. The participant took the notes mainly based on photographs and annotations with only few words. In addition, the note with minimum character size was written by another participant (1.84 cm) in the VR condition. This value indicated that writing in the presented VR classroom can be small. Overall, students are able to write many words and draw diagrams to explain a complex idea in VR.

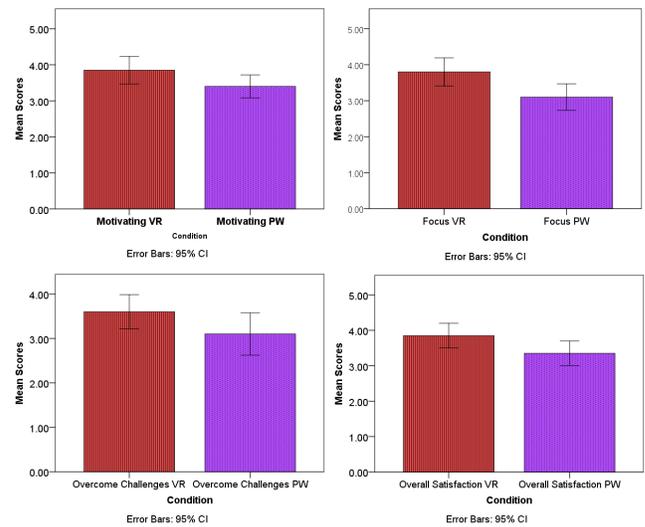

Figure 8: Results of mean Motivation scores (top left), mean Focus scores (top right), mean Overcome Challenges scores (bottom left), and mean Satisfaction scores (bottom right), between VR and PW conditions, respectively.

## 6.6 Qualitative Results

Most of the participants like the presented iVRNote interaction metaphor in the VR classroom experience. They mentioned that they could stay focused in VR than in PW. When studying video lectures by using a personal computer, they were often distracted by external sources. However, when they were in VR, what they could do were only listening to the instructors and taking notes. One participant reported, *"I may pause the video lecture and then start surfing on the Internet or cleaning my desk during the study. But I cannot do that in the VR classroom."* Several participants also pointed out that a new environment would help them stay focused. One participant said, *"The virtual classroom is new and attractive. I feel like I am more focused."*

More than half of the comments on the Wikipedia reference functionality in the VR classroom were positive. Several participants who preferred the function also suggested us to extend the search from Wikipedia to Google because in that way they could check more external sources. One participant said, *"I can obtain the information immediately by checking the Wikipedia. Besides, searching the word by voice is more efficient than searching the word by using a mobile phone."*

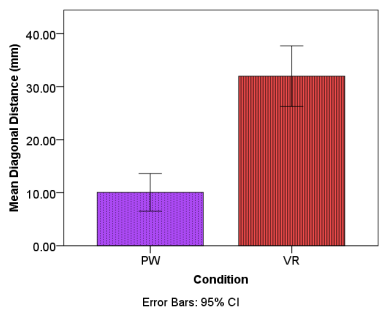

Figure 9: Results of mean diagonal distance of character writing between VR and PW condition.

Most of the participants liked the note taking system presented in the VR classroom. Particularly, they were grateful that the note taking by photograph greatly saved their effort in classes. One participant said, *"The note taking system is impressive. I like the photograph the most because it can be embedded in the note."* Another participant reported, *"Sometimes the professor would erase writings on blackboard or go to the next slide before I completed the notes. Taking a picture and then pasting it on the note can help prevent this problem."* However, in spite of the convenience, one commented, *"I used to take notes by writing although I can take a picture instead. Writing the important points improves my memory."* Besides the note taking by photograph, several participants were in favor of post-editing of notes as well. One participant said *"Sometime I have to organize the note by rewriting after the class. But with this, I don't need to do that."* As for the review, all of the participants commented that reviewing a specific part of the lecture by selecting notes is convenient. However, in the reviewing phase of the user study, we observed that the participants seldom used the review function. For the participants whose notes were informative and well-organized, they simply read their notes during the review session. For the remaining participants, they quickly went through the video lecture by dragging the time slider. After discussing with the participants about our observation, they explained that the lectures we chose in the user study were only about 30 minutes. They could review all topics in a lecture in 5 minutes. A participant said *"If I have to review the whole course taken in a semester, I would definitely use the function. There are too many topics to review, and I need an effective way."* Finally, since the tools, such as the pencil, eraser, and the utility knife, in the virtual environment have different shapes compared to the physical stylus, we suspected that the participants could have difficulties in reconciling the differences. However, the participants did not report this problem to us during the interview. They only complained that the eraser may occlude the area they would like to erase.

The participants pointed out several drawbacks and limitations of the presented VR classroom. Some of them were caused by the hardware. For example, they complained that the HMD is heavy and has low resolution. For the resolution problem, they had to experience the flickering artifacts of their notes. They also had to write characters in the virtual environment larger than those in the physical environment. Otherwise, they would have problems in reading them. Besides, the participants mentioned that the stylus tracking in the presented system could be improved. Sometimes the virtual stylus may shake or locate at the wrong position. We found that the problem usually occurs when the stylus is occluded by the users' hand or lies in the direction parallel to the viewing direction.

### 6.7 Discussions

We obtained the quantitative and qualitative feedback in the user study. Overall, the participants were very happy to be able to engage their attention, take notes, and review lectures using iVRNote interaction metaphor in a VR classroom. An interesting finding was that, the participants considered the presented system to be a very good tool for study in VR classrooms because disturbance from external sources were blocked. In the past, they were often distracted from the online courses and surfed on the Internet. But in the VR classroom they could only study because of the lack of distractions and the ability to focus. In addition, we measured the size of characters written by the participants. Although the participants tended to write larger characters in VR than in PW, the mean size of the characters in VR can be small at an average of 1.84 cm. The value indicates that the notes taken in VR can be complex. However, the participants also pointed out several drawbacks that make them less than willing to study in the VR classroom. For example, they felt that the HMD was heavy and of low resolution. They easily got tired in the study. In addition, they complained about the tracking accuracy, which would remind them what they saw was not real. However, despite of the disadvantages, the participants commented that the iVRNote was useful and would be more than willing to use it once the technical problems were solved.

## 7 CONCLUSIONS

We presented an interaction metaphor called iVRNote in a VR classroom for students in interactive note taking, learning, review and reflection in VR classroom situations when viewing lectures in pedagogical experiences. We conducted a user study evaluating the iVRNote interaction metaphor in a VR classroom learning against PW learning from viewing a lecture, study, and reflection. We found that participants can be immersed in the classroom and engage their full attention in study leveraging the interactive features of iVRNote. In addition, when they have problems in understanding materials in a lecture, they can check the Wikipedia to overcome difficulties. Considering that note taking is an important feature to facilitate an active learning experience in VR, we provided students with an intuitive interface to write down important points, well-organize the notes, and review lectures in a convenient and effective way through iVRNote. The results of our study revealed that the iVRNote had potential for greater pedagogical benefits, if VR displays and tracking systems gradually improve in the upcoming years. In future work, we aim to conduct an empirical evaluation examining participants' fine motor perception-action coordination in VR and PW writing, drawing, and sketching activities in pedagogical situations.

### ACKNOWLEDGMENTS

We would like to thank anonymous reviewers for their insightful comments. We are also grateful to all the participants who joined the user study. This work is partially supported by the Ministry of Science and Technology, Taiwan, under Grant No. 104-2221-E-009-041-MY3, 107-2811-E-009-004-, and 105-2221-E-009-135-MY3.